\documentclass[11pt]{article}
\usepackage[utf8]{inputenc}
\usepackage{amssymb, amstext, amscd, amsmath,color}
\usepackage{graphicx}
\usepackage{hyperref}
\usepackage{tabularx}
\usepackage{lscape}
\usepackage{subfig}
\usepackage{makecell}
\usepackage{matlab-prettifier}
\usepackage{mathtools}
\usepackage{geometry}
\usepackage{svg}
\graphicspath{{./figs/}}

\newtheorem{theorem}{Theorem}

\newtheorem{definition}{Definition}

\newtheorem{conj}{Conjecture}
\DeclareMathOperator{\E}{\mathbb{E}}
\DeclareMathOperator{\P1}{\mathbb{P}}

\DeclareMathOperator{\Var}{Var}
\newcommand\disteq{\mathrel{\overset{\makebox[0pt]{\mbox{\normalfont\tiny\sffamily $d$}}}{=}}} %command is \disteq, for independence equals

\title{Variance-Hawkes Process and its Application to Energy Markets}
\begin{document}
\maketitle
\begin{center}
Joshua McGillivray\\
University of Calgary\\
2500 University Drive NW, Calgary, AB, Canada, T2N 1N4\\
joshua.mcgillivray@ucalgary.ca\\

Anatoliy Swishchuk\\
University of Calgary\\
2500 University Drive NW, Calgary, AB, Canada, T2N 1N4\\
aswish@ucalgary.ca\\
\end{center}
\abstract{\noindent We define a new model using a Hawkes process as a subordinator in a standard Brownian motion. We demonstrate that this {\emph{Hawkes subordinated Brownian motion}} or more succinctly, {\emph{variance-Hawkes process}} can be fit to 2018 and 2019 natural gas and crude oil front-month futures log returns. This variance-Hawkes process allows financial models to easily have clustering effects encoded into their behaviour in a simple and tractable way. We also compare the simulations of a square of a variance Hawkes process with its Ito formula. We simulate both processes and compare their distributions, trajectories, and percent errors across multiple runs. We derive the generator relating to this Hawkes subordinated Brownian motion, calculate several moments, and conjecture its distribution. We also provide explicit solutions to the second moments of the Hawkes process and its intensity as well as the cross moment between the Hawkes process and its intensity in the case of an exponential kernel.\\} 

%%%%%%%%%%%%%%%%%%%%%%%%%%%%%%%%%%%%%%%
\section{Introduction}
Commodity spot prices have many unique and interesting properties that need to be reliably implemented in order to achieve an effective model. Some of these properties include volatility clustering, stochastic volatility, high kurtosis, volatility smile, jumps, and mean-reversion. Our aim in this paper is to directly address the first four properties in a novel and flexible way which allows researchers to easily add them to their models. Such models will be simple and tractable enough to implement in a real-world setting.

In any discussion about commodities, we must address the largest commodity market in the world: the energy sector. The energy sector contains two commodities that dominate every other commodity combined. Crude oil and natural gas are the largest two contributors to two well-known commodity indices: S\&P GSCI and BCOM. The S\&P GSCI tracks 24 commodities based on world production over the last 5 years. In 2007, the weight of crude oil was 51.30\% and 6.71\% from natural gas. for BCOM, the index tracks 19 commodities and is calculated using liquidity and world production. The weighting of crude oil was 13.88\% and natural gas was 11.03\% in 2007 \cite{Swishchuk2010}. In 2020 the S\&P GSCI set the reference percentage sector weight of the energy sector to 61.71\%. The next largest weighting was agriculture at 15.89\% \cite{SDJI2020}. Therefore we chose to focus on natural gas and crude oil for this paper. Specifically, we will be looking at 2018 and 2019 crude oil and natural gas front-month futures spot prices. 

The Hawkes process is useful because it is a simple tractable process that allows for clustering behaviour \cite{cui2020elementary}. This process has found use everywhere in the natural sciences from retweet dynamics on twitter to earthquake modeling, to of course financial modeling. Since financial markets share price information readily, they have lots of feedback, leading to volatility clustering on a regular basis. Examples of this behaviour are ubiquitous (See for example the flash crash of 2010, or Figures \ref{TS1}--\ref{TS4}). 

The Hawkes process has seen increasing use in financial spot price modeling both on a daily \cite{bacry2015hawkes}, \cite{bormetti2015modelling}, \cite{Swishchuk2018risk}, and an intra-day scale \cite{favetto2019european}. There has also been work on the application of Hawkes processes in limit order books (see \cite{swishchuk2019compound} for example). In 2015 Bacry et al. published an overview of recent literature on the Hawkes process, this paper is often used as an entry point for those interested in the Hawkes process \cite{bacry2015hawkes}. In the same year, Bormetti et al. showed that the Hawkes process can robustly model financial assets after accounting for the possibility of multiple jumps at once \cite{bormetti2015modelling}. Also in 2015, Laub et al. published an overview paper that summarized the background, historical developments, and major aspects of the Hawkes process \cite{laub2015hawkes}. In 2018, Swishchuk et al. derived a risk model based on a compound Hawkes process \cite{Swishchuk2018risk}. In 2019, Favetto showed that the European intra-day electricity market follows a Hawkes process using Q-Q analysis \cite{favetto2019european}. In the same year, Swishchuk et al. applied the compound Hawkes process to limit order books and derived a few results relating to the law of large numbers and the central limit theorem \cite{swishchuk2019compound}. 

Subordination on the other hand has seen an equal if not greater amount of success than the Hawkes process. Two of the most well-known subordinator processes are the normal inverse Gaussian (NIG), and the variance-gamma (VG) processes. Both of these processes are common in financial modeling. The NIG model was introduced into the financial literature in 1997 by Barndorff \cite{barndorff1997normal}. The VG process has also seen much use in financial contexts. It was first introduced in 1990 by Madan et al. \cite{madan1990variance}. In 2004 Carr et al. used time changed Levy processes to address issues with asset prices jumps, stochastic volatility, and correlated volatilities and returns \cite{carr2004time}. In 2017 Borokova et al. showed that the subordination of a general Levy process can be used reliably to model electricity as a time-changed model. Their model performed quite well. They go into detail about calendar versus business time, stating that business time changes as a function of activity, and thus a time-change approach can capture this nonlinear time phenomenon \cite{borovkova2017electricity}. 

Curiously missing from all of the aforementioned subordination sources is the application of time change to pure jump processes. Although the gamma process is a pure jump process, the authors of the variance-gamma process do not go into depth on the interpretation of this as a time change. The literature on subordination appears to fall into two main categories: (1) research following a specific choice of continuous subordinator (e.g. the gamma process or inverse gaussian process); and (2) research into subordinators as fully general Levy processes. This leaves a gap in the study of subordinators where pure jump processes are neglected. This is because subordinators typically are interpreted as a transformation in time, but there is no mathematical restriction that mandates this interpretation provided the subordinator of choice is nondecreasing. Indeed this view is only a special case of the work done by Carr et al. and Borovkova et al.. In this paper, we show there is value in a standard Brownian motion with Hawkes subordinator. We call this process the {\emph{variance-Hawkes process}}. It is tractable, simple, has few parameters, and can be easily added to existing models to grant them jump behaviour, stochastic volatility, and volatility clustering.

%Outline
This paper is outlined as follows: section 2 defines a {\emph{variance Hawkes process}} and gives a few relevant facts about the Hawkes process. Section 3 describes why the variance Hawkes process is unique, relevant, and includes a small discussion on why discontinuous subordinators are worthy of attention. In section 4 we present evidence for the variance Hawkes process in natural gas and crude oil markets. In section 5 we derive the generator of a variance Hawkes process. In section 6 we derive the Ito formula for the variance of a variance Hawkes process and perform several simulations of this process.
%%%%%%%%%%%%%%%%%%%%%%%%%%%%%%%%%%%%%%%
\section{The Hawkes Process}\label{BNt}
Define the pair $(N_t,\lambda_t)$ to be a {\emph {simple Hawkes process with exponential kernel}} where
\[\lambda_t=v+\alpha\sum_{T_r<t}e^{-\beta(t-T_r)}\label{int}\]
\noindent is the intensity function and $T_r$ is the arrival times of the jumps. $\alpha, \beta, v>0$ and $\alpha\neq\beta$. Typically we also want $\alpha<\beta$ to prevent a runaway feedback loop of jumps, but in our cases, this may not be necessary. Note that $N_t$ will sometimes also be referred to as a Hawkes process depending on the context.

It is well known that the Hawkes process satisfies for small $\Delta>0$,
\begin{align}
N_0=0,\\
\P1(N_{t+\Delta}-N_t\geq2)&=o(\Delta),\label{p2}\\
\P1(N_{t+\Delta}-N_t=1)&=\lambda_t\Delta+o(\Delta),\label{p1}\\
\P1(N_{t+\Delta}-N_t=0)&=1-\lambda_t\Delta+o(\Delta).\label{p0}
\end{align}
If $X$ is a random variable then $X=o(\Delta)$ is taken to mean 
\[\lim_{\Delta\downarrow 0}\frac{X}{\Delta}=0.\]

We will need two properties of the simple Hawkes process given by Cui et al.~\cite{cui2020elementary}:
\begin{align}
\E(N_t)=\frac{vt}{1-\alpha / \beta}-\frac{\alpha / \beta}{(1-\alpha / \beta)^2}\frac v\beta\left(1-e^{-\beta(1-\alpha/\beta)t}\right),\label{mean}
\end{align}
\begin{align}
\E(\lambda_t)=\frac{v}{\beta-\alpha}\left(\beta-\alpha e^{-(\beta-\alpha)t}\right).\label{mean2}
\end{align}
Note that $\int_0^t\E(\lambda_t)dt=\E(N_t)$.

Before we discuss clustering models, it will be valuable to derive some properties of the Hawkes process, which will be used as a driver for the clustering behaviour. Cui et al.~\cite{cui2020elementary} provide a method and a formula for the moments of the Hawkes process with the exponential kernel. Their results may be used for the expected values of any well-behaved function $f(x)$ of the Hawkes process, $\E(f(x))$, given in the form of a differential equation. The formula which is most useful for the process's moments is given by:
\begin{align}
\frac{d}{dt}\E(N^m_t\lambda^n_t)&=n\beta v\E(N^m_t\lambda^{n-1}_t)-n\beta \E(N^m_t\lambda^n_t)\nonumber\\
&+\sum_{j=0}^{m-1}\binom{m}{j}\E(N^j_t\lambda^{n+1}_t)\nonumber\\
&+\sum_{j=0}^m\sum^{n-1}_{i=0}\binom{m}{j}\binom{n}{i}\alpha^{n-i}\E(N^j_t\lambda^{i+1}_t).\label{mom}
\end{align}
By setting $n=0$ and $m=2$, we can use this formula to obtain a differential equation in terms of the second moment. Noting that the first, second, and fourth terms above are 0, the differential equation for the second moment is given using this formula by:
\begin{equation}
\frac{d}{dt}\E(N_t^2)=\E(\lambda_t)+2\E(N_t\lambda_t),
\end{equation}
which requires $\E(N_t\lambda_t)$. This can be derived using equation \eqref{mom} by setting $n=m=1$ and solving the resulting differential equation.
\begin{equation}
\frac{d}{dt}\E(N_t\lambda_t)=\beta v\E(N_t)+(\alpha-\beta)\E(N_t\lambda_t)+\E(\lambda_t^2)+\alpha \E(\lambda_t)
\end{equation}
requires $\E(\lambda_t^2)$, this can be derived using equation \eqref{mom} by setting $n=2, m=1$ and solving the resulting differential equation.
\begin{equation}
\frac{d}{dt}\E(\lambda_t^2)=2\beta v\E(\lambda_t)-2\beta \E(\lambda_t^2).
\end{equation}
Therefore, to determine the second moment of the Hawkes process, we need to solve three first-order linear differential equations. These equations are first-order non-homogenous linear differential equations which have a solution easily determined by the variation of parameters method. This work is tedious but not difficult, Cui et al. \cite{cui2020elementary} have already derived some of these values for a simple Hawkes process. Since our setting is applied and we will calibrate $\alpha$ and $\beta$ using real-world data, we assume that $\alpha\neq\beta$, though we note that the $\alpha=\beta$ case is given in their paper. Explicit expressions for these three moments are given in Appendix \ref{explicitsol}.
%%%%%%%%%%%%%%%%%%%%%%%%%%%%%%%%%%
\section{Subordination and a Clustering Model Framework}\label{SubordAndCluster}
Subordination and time change models have been used widely for decades and have all found at least one niche in the financial sector. In this section, we propose a new model using a foundational approach to the understanding of how a market price moves. On the open market, the price of an asset is its most recent sale price. Naturally, this price will be updated once a new sale occurs. Therefore the price of an asset on the market is not directly time dependent, rather it depends on the arrival times of orders. This implies a subordination approach may be a more appropriate direction for price modeling. We note that this approach is also referred to in the literature as a conversion from calendar time to \emph{business time}. Borokova et al. \cite{borovkova2017electricity} have used business time to great success. In their paper, they consider two business time transformations: subordination and absolutely continuous time change. In this section, we avoid the term ``business time'' since our subordinator of choice is not continuous. This choice of discontinuous subordinator is justified because trade order arrivals are inherently discontinuous. We will elaborate on this idea shortly.

Favetto in \cite{favetto2019european} discovered the arrival times of orders in European intra-day electricity markets appear to follow a Hawkes process using Q-Q analysis. We can apply the same methodology to North American markets. If the arrival times of orders follow a Hawkes process, then it is possible that price spikes in these markets also follow a Hawkes process since a rush of buy or sell orders is likely to cause volatility. We assessed the arrival times of orders and saw their intra-day arrival times exhibited clustering (see Figures \ref{QQCO}--\ref{VolNG}). In this case, a model of the form $e^{S(t)}$ where $S(t)=\varepsilon(t)+L(N_t)$ may be effective in price modeling, where $\varepsilon(t)$ is some stochastic process (for example a geometric Brownian motion or Ornstein--Uhlenbeck process), $L(t)$ is a Levy process, and $N_t$ is a Hawkes process with the exponential kernel. We will call models of this form \emph{clustering models} since a self-exciting jump process as the subordinator will lead to clusters of volatility. For example if
\[dS(t)=\kappa(\theta-S(t))dt+\sigma dW_t+d(L(N(t))),\]
we would refer to this process as a \emph{clustered Ornstein--Uhlenbeck process}. Or perhaps one may consider an \emph{Ornstein--Uhlenbeck process with stochastic mean reverting level}:
\[dS(t)=\kappa(L(N(t))-S(t))dt+\sigma dW_t.\]
This technique awards a simple and tractable way to encode volatility clustering into already well-established models. The latter would be able to account for stochastic changes in the mean-reverting level, which may cluster based on fluxes in demand associated with market shocks. This kind of behaviour is exhibited in the time series of natural gas futures spot prices in Figure \ref{TS2} and Figure \ref{TS4}.

To our knowledge, the literature on subordination appears to fall into two main categories: (1) research following a specific choice of continuous subordinator (e.g. the gamma process or inverse Gaussian process); and (2) research into subordinators as fully general Levy processes (see for example \cite{borovkova2017electricity}). This leaves a gap in the study of subordinators where pure jump processes are neglected. This is because subordinators typically are interpreted as a transformation in time, but there is no mathematical restriction that mandates this interpretation provided the subordinator of choice is nondecreasing. We may interpret discontinuous subordinators as counting the number of order arrivals in an interval, and the model as inferring the price from those arrivals. Although it is tempting to view the index of a jump-diffusion process as a continuous time parameter, there are no assumptions that require this. If these processes `progress' based on trade arrivals, then the price processes experience `time' with the arrivals of trades, which are necessarily discontinuous. This means other subordination models with continuous subordinators are approximating a discontinuous process with a continuous one. There is naturally nothing wrong with this sort of approximation, but if the arrival time distribution is known then a continuous subordinator may not be the best choice. The adherence to continuous time under this perspective may be unnecessary and potentially distracting. Success with this approach would shift the focus of price modelling from attempting to model the distribution of time series to attempting to model the distribution of arrival times of trades.

\section{The Variance-Hawkes Process and Its Justification in Energy markets}

\subsection{Definition of the Variance-Hawke Process}

Define $B(t)$ as an independent standard Brownian motion, i.e. $B(t)$ has mean 0 and variance $t$. Define a {\emph{Hawkes subordinated Brownian motion}} or more succinctly, a {\emph{variance-Hawkes process}} to be the composition of the Hawkes process with the Brownian motion: $B(N_t)$. We arrive at our definition of a variance Hawkes process:

\begin{definition}[{\bf{Variance-Hawkes Process}}]
A ``variance-Hawkes process'' is a process formed by the subordination of a Hawkes process into a standard Brownian motion:
 \[\mathcal{V}_t:=B(N_t).\]
\end{definition}

%%%%%%%%%Figure 5

\begin{figure}[h!]
\centering
\subfloat{\scalebox{1}{\includegraphics[width=0.9\linewidth]{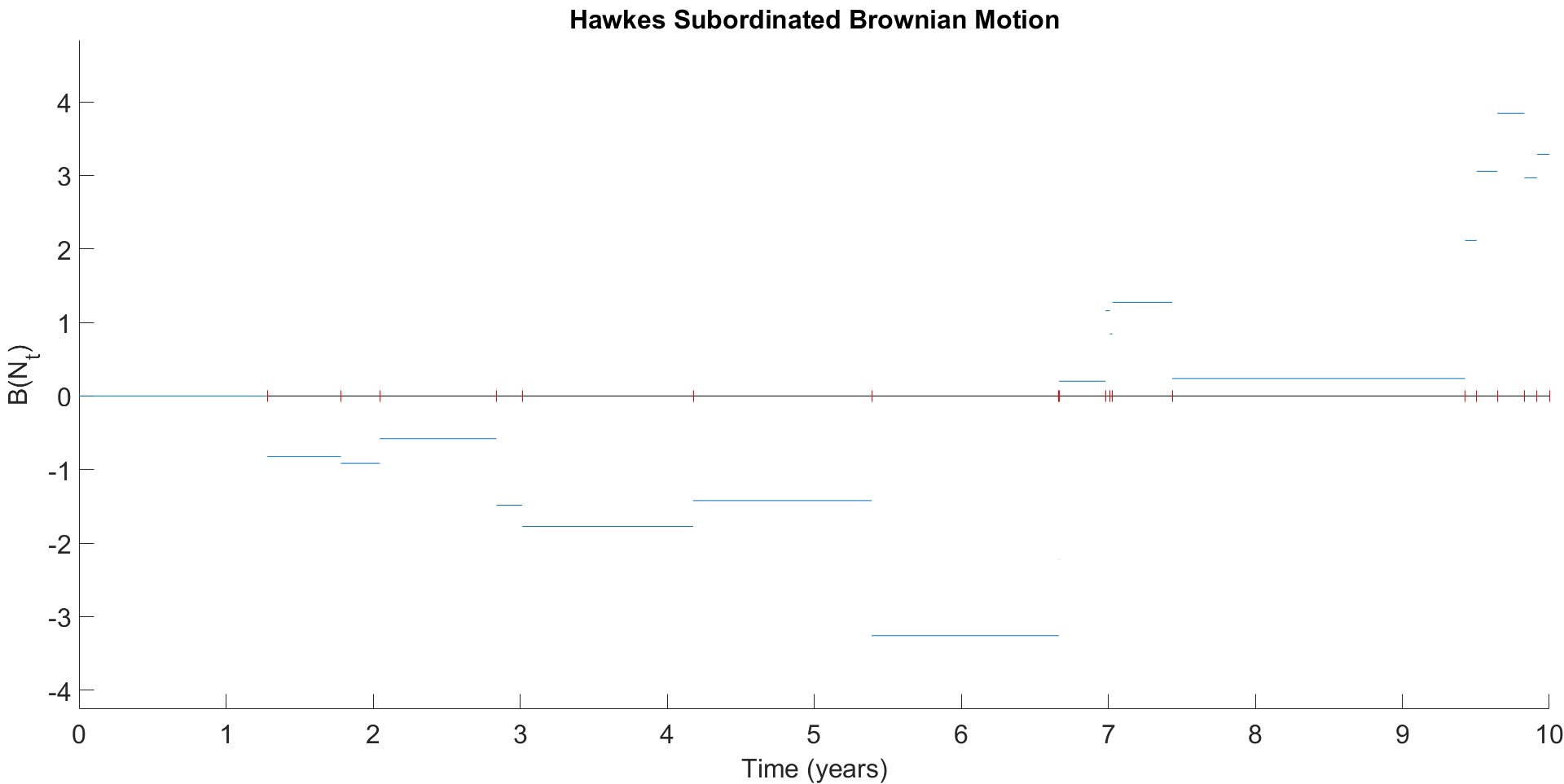}}}
\caption{A standard Brownian motion with simple Hawkes subordinator (variance-Hawkes process): $v=1$, $\alpha=1$, and $\beta=2$; vertical red lines on the horizontal axis denote jump times.}\label{bnt1}%
\end{figure}
The latter camp of research (see sec. 3) is still applicable to discontinuous subordinators since Levy processes include some jump processes. However, we need to be careful since not all jump-diffusion processes are Levy processes. In the case of the Hawkes process, the requirement to become a Levy process restricts us to an exponential kernel ($\phi(t)=X_t e^{-\beta t}$, where $X_t$ is a Markov process \cite{bacry2015hawkes}). Nonetheless, these models are interesting and have complex behaviour relative to their number of parameters. For example, if we were to consider a variance Hawkes process i.e. a standard Brownian motion with simple Hawkes subordinator, $N_t$ (and $\phi(t)=\alpha e^{-\beta t}$; $\alpha,\beta>0$) which has the form $B(N_t)$, it has only three parameters: $\alpha$, $v$, and $\beta$ but exhibits stochastic volatility and clustering.  Figure \ref{bnt1} is a simulation of $B(N_t)$.

%%%%%%%%%%%%%%%%%%%%%%%%%%%%%%%%%%%%%%%%%%%

\subsection{Initial Evidence for $B(N_t)$ in Crude Oil and Natural Gas Markets}

We may calibrate the $\alpha$, $\beta$, $v_0$, and $v$ to see if $B(N_t)$ is a good fit for 2018 and 2019 crude oil and natural gas front-month futures, where $v_0$ is the initial intensity of the Hawkes process. Since we have not yet derived and vetted calibration techniques for $B(N_t)$, we chose to fit these datasets by hand using trial and error. Taking $B(N_t)$ alone as a price model would be akin to using the Bachelier model as a price model, thus we took the log returns of our dataset and hand fit $B(N_t)$ since most price models today use an exponential form. Additionally, $B(N_t)$ is not intended to model prices outright, rather it is intended as a tool that can be easily added onto other models to improve their performances. In the interest of illustration and to minimize the number of parameters we need to balance, we took a relatively simple model:
\[\ln\left(\frac{S_{t+1}}{S_t}\right)=a+b\hat\sigma B(N_t),\]
where $\hat\sigma$ is the sample standard deviation of the dataset, $a$ is some real constant, and $S_t$ is the asset price.

The results are given by Figures \ref{2018NGBNt}--\ref{2019COBNt}. The chosen fits are not as precise as they could be had they have been chosen mathematically. Nonetheless, the distributions of the various $B(N_t)$ models demonstrate good overlap with all four experimental datasets despite the basic calibration and model structure. In particular the tail behaviour of Figures \ref{2019NGBNt}, \ref{2018COBNt}, and \ref{2019COBNt}, appear to match the experimental distributions quite closely. One can imagine a more well-vetted model such as the Schwartz one-factor model modified with $B(N_t)$ would perform substantially better. The model chosen in this paper does not even contain mean reversion properties, thus much better fits for models using $B(N_t)$ are certainly possible. In fact we are not restricted to only the $B(N_t)$ framework, we can easily extend this behaviour to a gaussian process:
\[G(N_t)=\mu +\sigma B(N_t),\]
where $\mu\in\mathbb{R}$ and $\sigma>0$ are constants. Clearly from here $\mathbb{E}(G(N_t))$ and $\text{Var}(G(N_t))$ are easily accessible using only the properties of the expected value and variance. Similarly, further extensions are clearly possible, for example into Levy processes.

%%%%%%%%%%%%%
\subsection{Simulated clustered Gaussian process with 2018/2019 natural gas and crude oil futures log-returns}

\begin{figure}[h!]
\centering
\subfloat{\scalebox{0.6}{\includegraphics[width=0.9\linewidth]{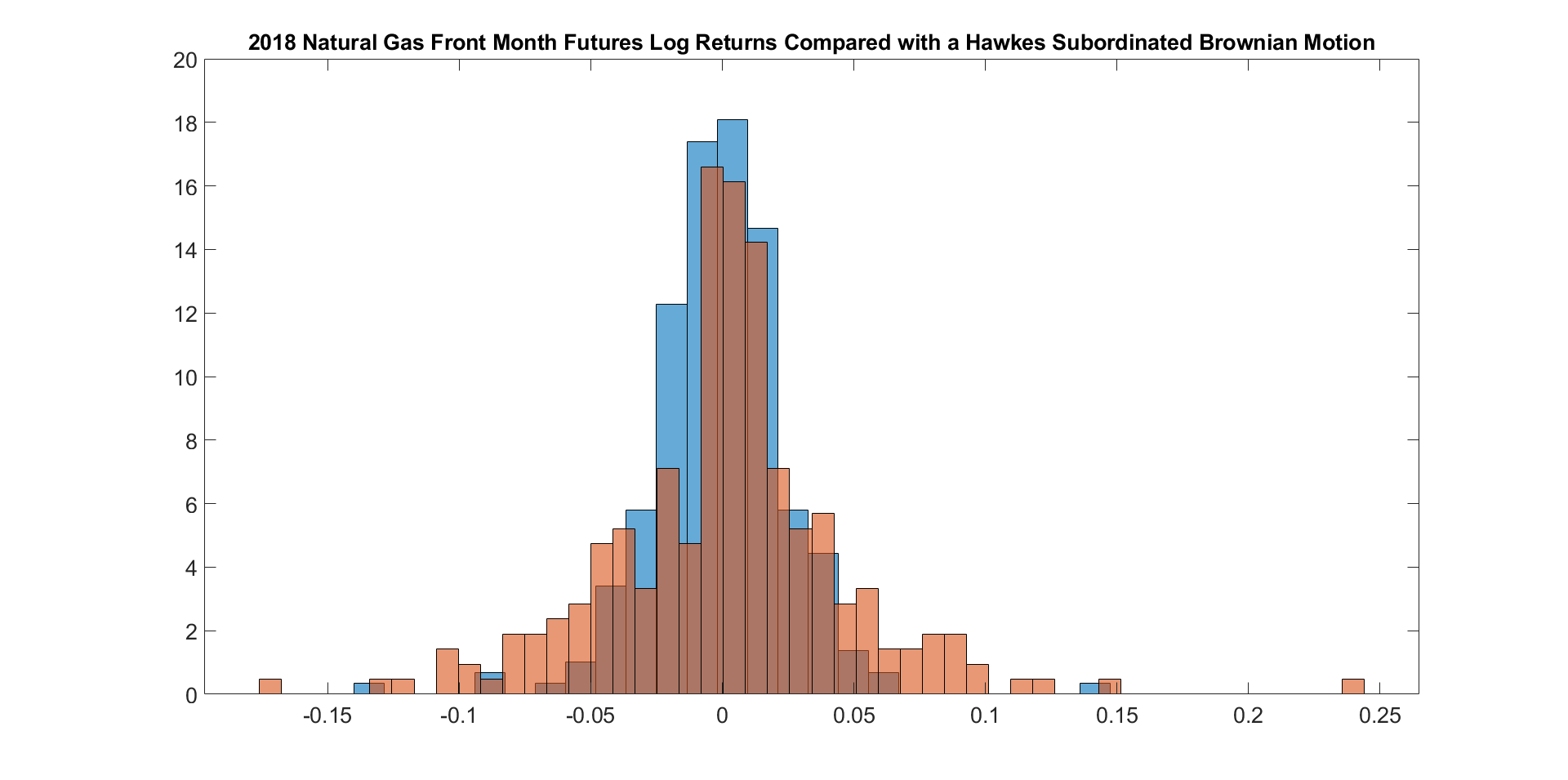}}}
\caption{The distribution of log returns of one year of 2018 front-month natural gas futures prices (blue) and the distribution of a simulated Hawkes subordinated Brownian motion (red) with parameters $v_0=401$, $v=400$, $\beta=800$, $\alpha=700$, $T=1$, $a=0$, and $\hat\sigma=0.0320$.}\label{2018NGBNt}%
\end{figure}

\begin{figure}[h!]
\centering
\subfloat{\scalebox{0.6}{\includegraphics[width=0.9\linewidth]{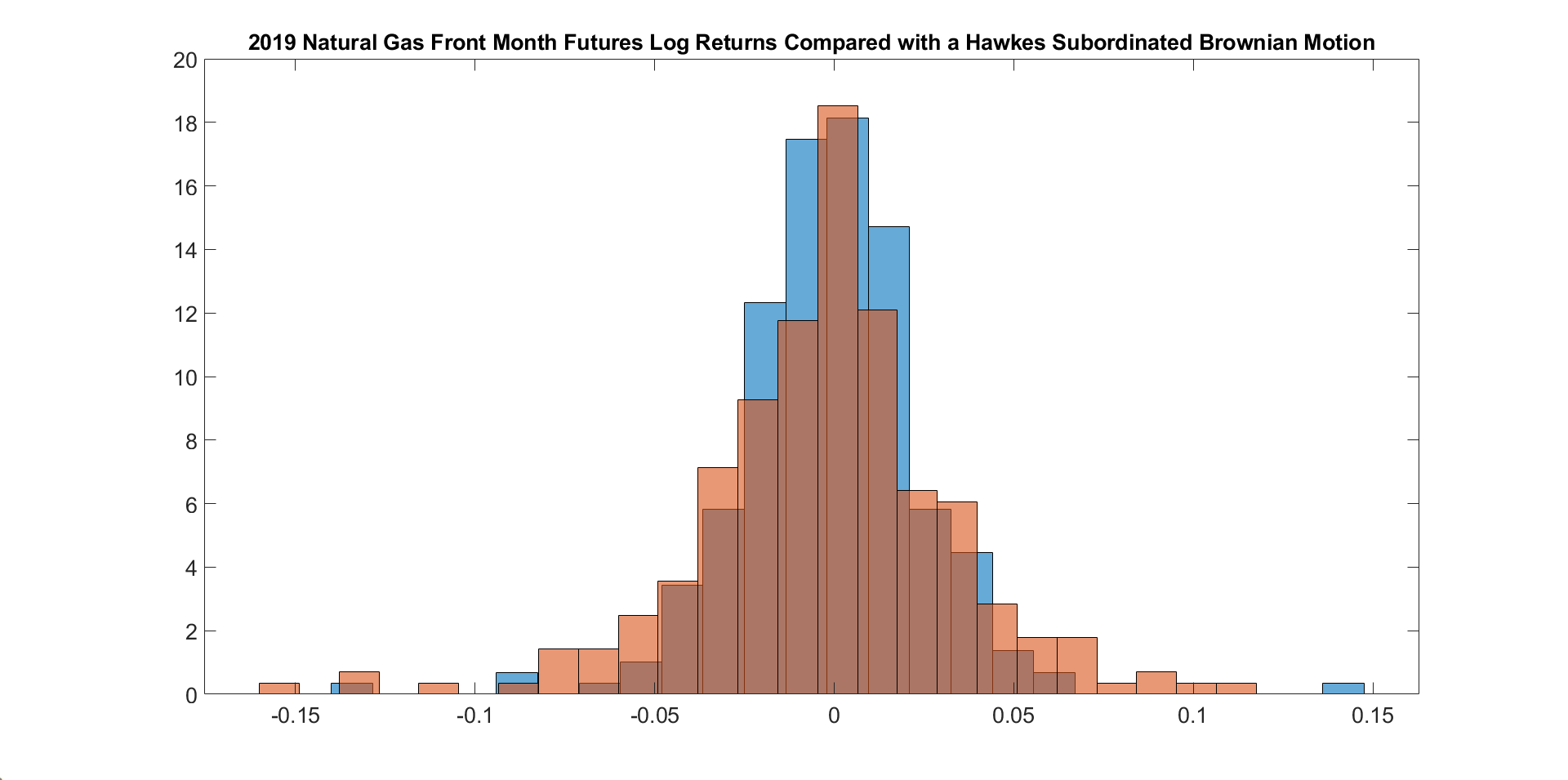}}}
\caption{The distribution of log returns of one year of 2019 front-month natural gas futures prices (blue) and the distribution of a simulated Hawkes subordinated Brownian motion (red) with parameters $v_0=451$, $v=450$, $\beta=800$, $\alpha=700$, $T=1$, $a=0$, and $\hat\sigma=0.0267$.}\label{2019NGBNt}%
\end{figure}

\begin{figure}[h!]
\centering
\subfloat{\scalebox{0.6}{\includegraphics[width=0.9\linewidth]{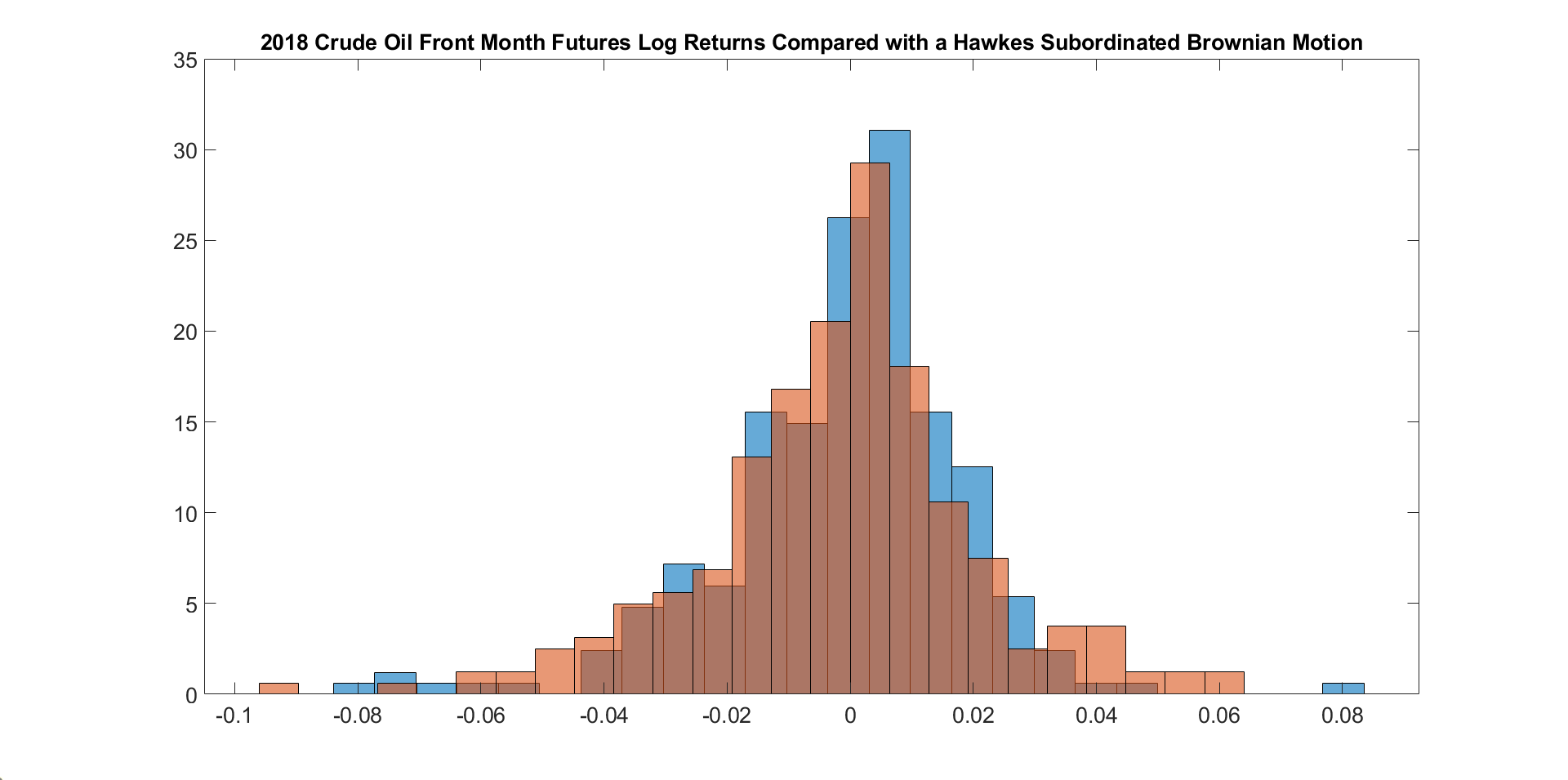}}}
\caption{The distribution of log returns of one year of 2018 front-month crude oil futures prices (blue) and the distribution of a simulated Hawkes subordinated Brownian motion (red) with parameters $v_0=401$, $v=400$, $\beta=800$, $\alpha=700$, $T=1$, $a=0$, and $\hat\sigma=0.0212$.}\label{2018COBNt}%
\end{figure}

\begin{figure}[h!]
\centering
\subfloat{\scalebox{0.6}{\includegraphics[width=0.9\linewidth]{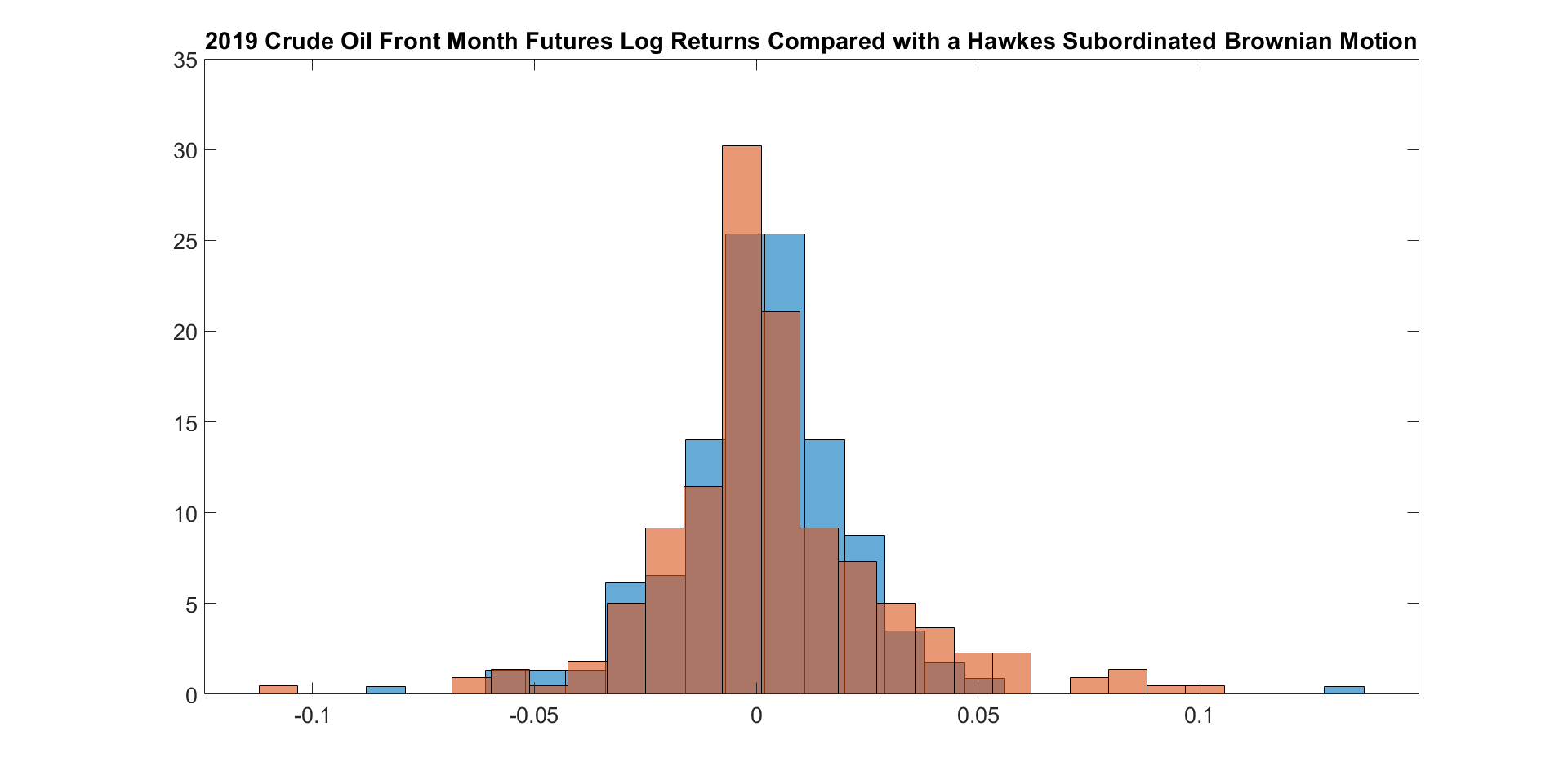}}}
\caption{The distribution of log returns of one year of 2019 front-month crude oil futures prices (blue) and the distribution of a simulated Hawkes subordinated Brownian motion (red) with parameters  $v_0=401$, $v=400$, $\beta=800$, $\alpha=700$, $T=1$, $a=0$ and $\hat\sigma=0.0267$.}\label{2019COBNt}%
\end{figure}

Figures \ref{2018NGBNt}--\ref{2019COBNt} compare the {\emph{clustered Gaussian process}} (red) given by $a$ with the log returns of 2018 and 2019 natural gas and crude oil front month futures (blue). The parameters were decided via hand calibration, trial, and error since a calibration method has not yet been vetted. Given the simplicity of the model, the fit is remarkably good. In all cases $b=\frac{1}{e}$ was chosen where $e$ is Euler's constant.

%%%%%%%%%%%%Figures 7-10 and 14-17
%%%%%%%%%%

\newpage
%

%%%%%%%%%%%%%%Fig 14-15 and 16-17

\subsection{Clustering behaviour in crude oil and natural gas futures trading volume}
\begin{figure}[h!]
\centering
\subfloat{\scalebox{0.6}{\includegraphics[width=0.9\linewidth]{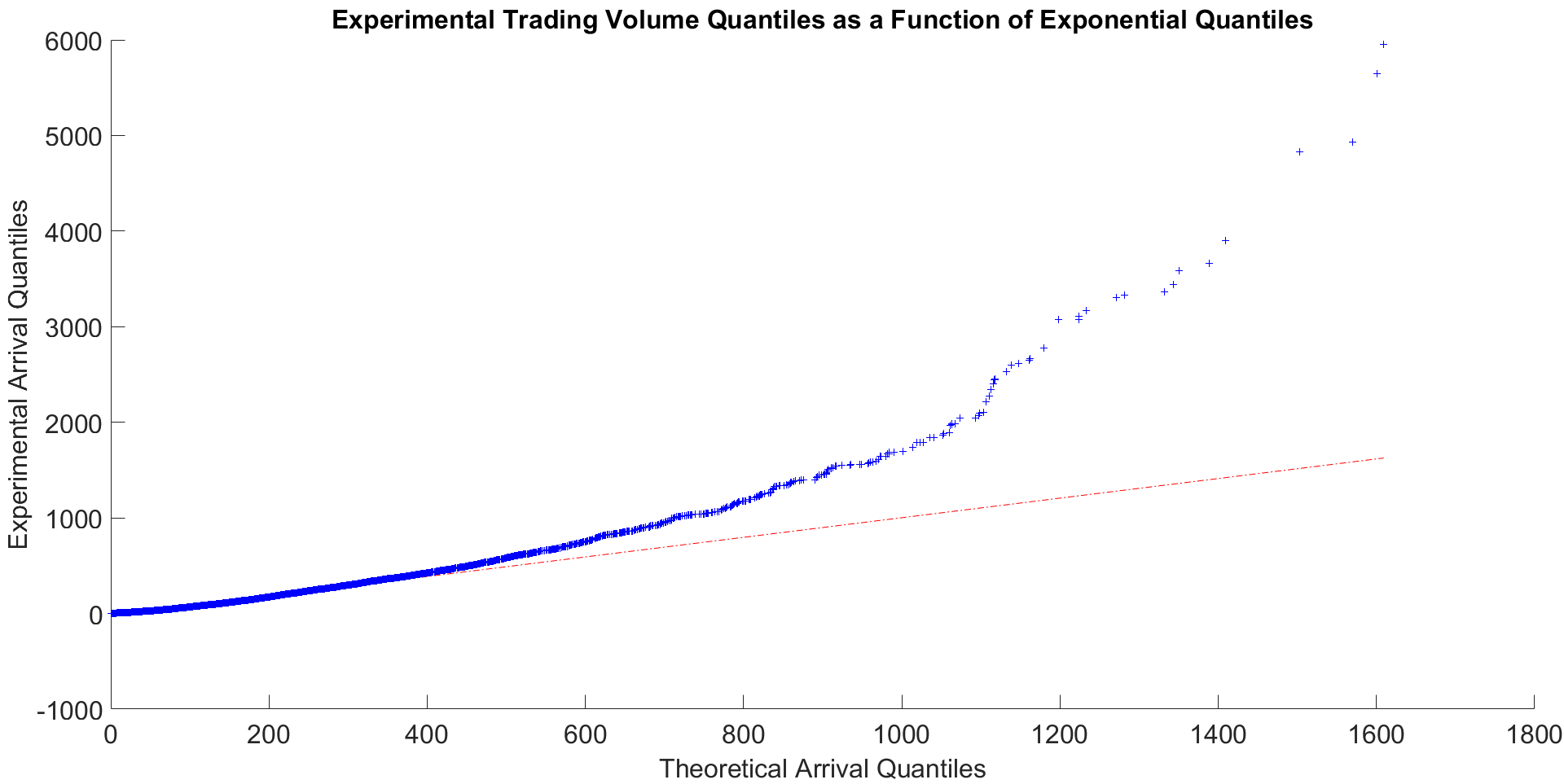}}}
\caption{2024 WTI crude oil futures trade volume quantiles as a function of simulated exponential quantiles.}\label{QQCO}%
\end{figure}

\begin{figure}[h!]
\centering
\subfloat{\scalebox{0.6}{\includegraphics[width=0.9\linewidth]{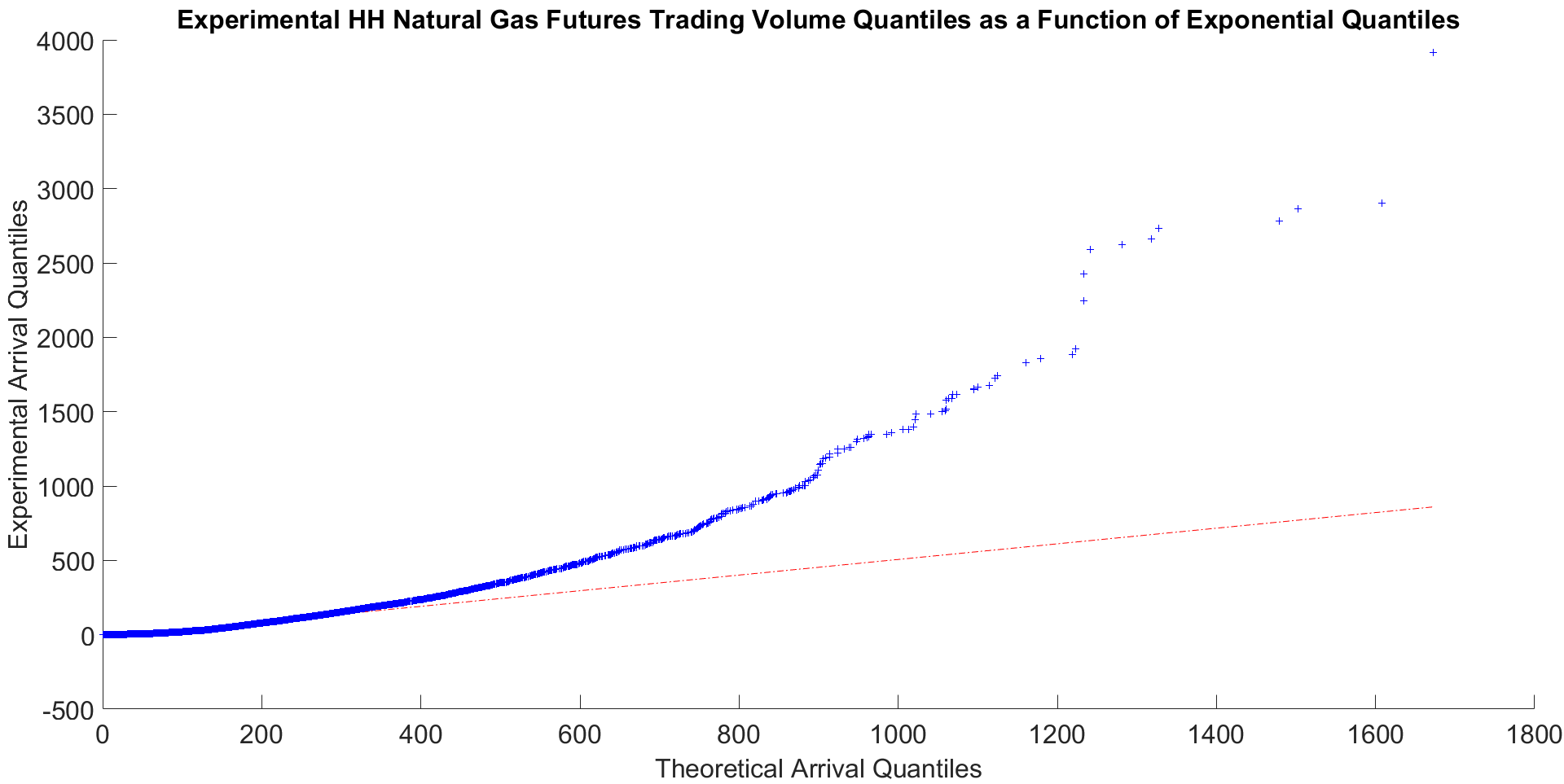}}}
\caption{2024 HH natural gas futures trade volume quantiles as a function of simulated exponential quantiles.}\label{QQNG}%
\end{figure}

Figures \ref{QQCO} and \ref{QQNG} show that the per-minute trade volume quantiles of WTI crude oil and NYMEX natural gas futures do not follow exponential quantiles. This demonstrates that trades occur more often than expected under a Poisson model. A Hawkes model may be able to explain this discrepancy through its clustering effects.
\newpage
\subsection{Trade volume clustering in crude oil and natural gas trading volumes}
\begin{figure}[h!]
\centering
\subfloat{\scalebox{0.6}{\includegraphics[width=0.9\linewidth]{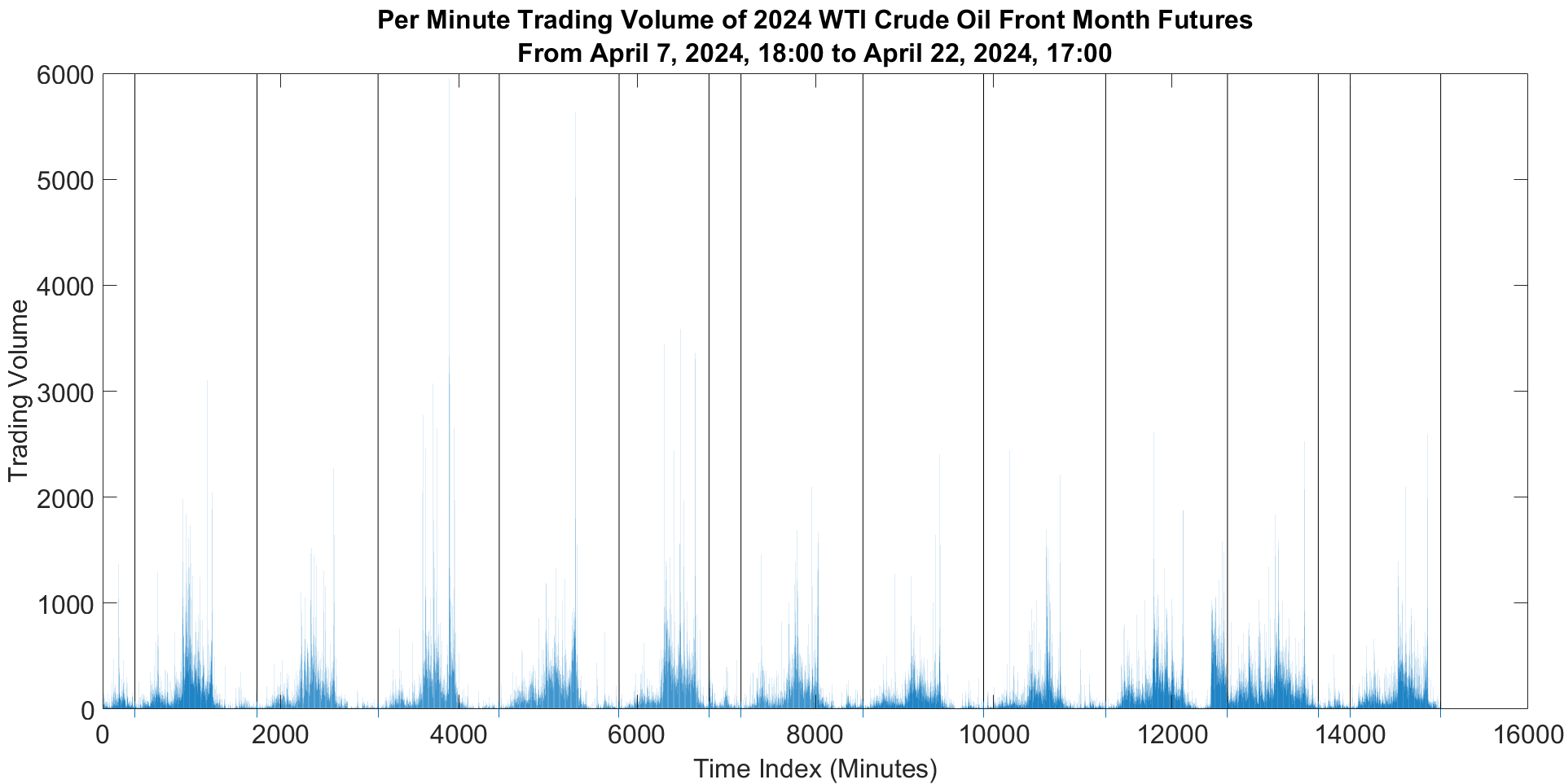}}}
\caption{2024 WTI crude oil futures trade volume as a function of time in minutes; vertical black lines denote the end of each day in the dataset.}\label{VolCO}%
\end{figure}

\begin{figure}[h!]
\centering
\subfloat{\scalebox{0.6}{\includegraphics[width=0.9\linewidth]{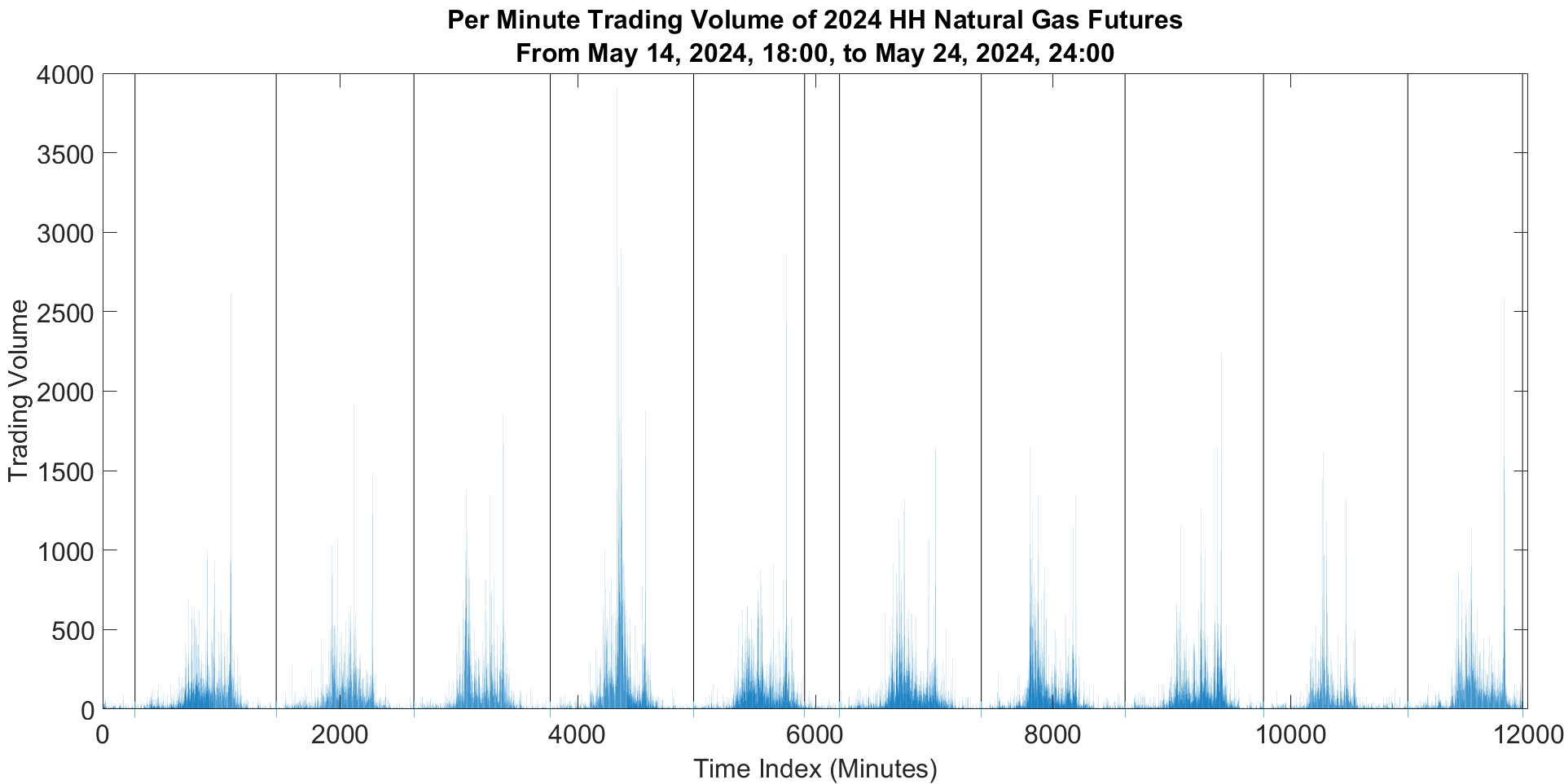}}}
\caption{2024 HH natural gas futures trade volume as a function of time in minutes; vertical black lines denote the end of each day in the dataset.}\label{VolNG}%
\end{figure}

Figures \ref{VolCO} and \ref{VolNG} show that trade volume is not uniformly distributed throughout the trading day. It appears trades are clustered towards the midday and late afternoon, this behaviour would not be expected under a Poisson model. These figures serve as evidence for trade clustering which lends towards a Hawkes model for trade arrival times.
\clearpage

%%%%%%%%%%%%%%%%%%%%%%%%%%%%%%%%%%
\section{Some Properties of Variance-Hawkes Process}

\subsection{Generator of a Variance-Hawkes Process}

Define $Y_t=(\lambda_t,N_t,B(N_t))$ where $N_t$ is a simple Hawkes process with exponential kernel and intensity $\lambda_t$, and $B_t$ is an independent standard Brownian motion. The generator for this process is given for $t>0$ by
\begin{equation}
\mathcal{A}f=\lim_{t\downarrow 0}\frac{T_tf-f}{t}\label{generator}
\end{equation}
where $f\in\left\{ f \in C_0(E):\lim_{t\downarrow 0}\frac{T_tf-f}{t}\text{ exists as a uniform limit}\right\}$, $C_0(E)$ denotes the Banach space of continuous functions on E which vanish at infinity with the supremum norm, and $T_tf(y)=\E(f(Y_t)|Y_0=y)$. Thus we need to calculate $\E(f(Y_t)|Y_0=y)$. We will impose one additional assumption on $f(l,n,b)$: all of its partial derivatives up to the second order exist and are finite. Note that the input parameters $l,n,b$ are chosen to invoke the notion of $\lambda_t, N_t,$ and $B(N_t)$ respectively. Note that $\lim_{t\downarrow 0}o(t^2)/t=0$.

Observe that we can approximate our expectation with $A_2(Y_t)$, a second-order trivariate Taylor approximation centered at $y=(\lambda_0, N_0, B(N_0))$: $\E(f(Y_t)|Y_0=y)=\E(A_2(Y_t)|Y_0=y)+o(t^2)$. Substituting this into equation \eqref{generator} we have:
\begin{equation}
\mathcal{A}f(y)=\lim_{t\downarrow 0}\frac{\E(A_2(Y_t)|Y_0=y)-f(Y_t)}{t}+\lim_{t\downarrow 0}\frac{o(t^2)}{t}=\lim_{t\downarrow 0}\frac{\E(A_2(Y_t)|Y_0=y)-f(Y_t)}{t}.
\end{equation}
Therefore, we must calculate $\E(A_2(Y_t)|Y_0=y)$. The general form for a second-order trivariate Taylor approximation centered at $y=(\lambda_0, N_0, B(N_0))$ is
\begin{align}
A_2(l,n,b)&=f(y)+\frac{\partial f(y)}{\partial l}(l-\lambda_0)+\frac{\partial f(y)}{\partial n}(n-N_0)+\frac{\partial f(y)}{\partial b}(b-B(N_0))\nonumber\\
&\ \ +\frac{\partial^2f(y)}{\partial l\partial n}(l-\lambda_0)(n-N_0)+\frac{\partial^2f(y)}{\partial l\partial b}(l-\lambda_0)(b-B(N_0))\nonumber\\
&\ \ +\frac{\partial^2f(y)}{\partial n\partial b}(n-N_0)(b-B(N_0))+\frac12\frac{\partial^2f(y)}{\partial l^2}(l-\lambda_0)^2\nonumber\\
&\ \ +\frac12\frac{\partial^2f(y)}{\partial n^2}(n-N_0)^2+\frac12\frac{\partial^2f(y)}{\partial b^2}(b-B(N_0))^2.
\end{align}
Next, observe that $(\lambda_0, N_0, B(N_0))=(v,0,0)$ since $B(0)=N_0=0$. Applying this fact, evaluating the polynomial at $Y_t=(\lambda_t,N_t,B(N_t))$, and taking the expected value of both sides yields
\begin{align}
\E(A_2(l,n,b))&=f(y)+\frac{\partial f(y)}{\partial l}\E(\lambda_t-v)+\frac{\partial f(y)}{\partial n}\E(N_t)+\frac{\partial f(y)}{\partial b}\E(B(N(t)))\nonumber\\
&\ \ +\frac{\partial^2f(y)}{\partial l\partial n}\E[(\lambda_t-v)N_t]+\frac{\partial^2f(y)}{\partial l\partial b}\E[(\lambda_t-v)B(N_t)]\nonumber\\
&\ \ +\frac{\partial^2f(y)}{\partial n\partial b}\E[N_tB(N_t)]+\frac12\frac{\partial^2f(y)}{\partial l^2}\E[(\lambda_t-v)^2]\nonumber\\
&\ \ +\frac12\frac{\partial^2f(y)}{\partial n^2}\E[N_t^2]+\frac12\frac{\partial^2f(y)}{\partial b^2}\E[(B(N_t))^2].\label{genbnt}
\end{align}
This equation has nine unknowns to determine: $\E(\lambda_t)$, $\E(N_t)$, $\E(B(N_t))$, $\E(\lambda_tN_t)$, $\E(\lambda_tB(N_t))$, $\E(N_tB(N_t))$, $\E(\lambda^2_t)$, $\E(N_t^2)$, and $\E((B(N_t))^2)$. Equations \eqref{closedformEl2}--\eqref{closedformEN2} demonstrate we can already calculate five of these unknowns and since we have their exact forms, the upcoming related limits to these five unknowns will be calculable as well. Therefore we only need to find the expectations associated with $B(N_t)$. Since $B(N_t)$ is an interwoven process with standard normal-sized jump heights at Hawkes distributed jump times, we know that $\E(B(N_t))=0$ by symmetry. The second moment will be substantially more work but should be attainable via conditioning. $B(N(t))$ is not a Markov process, so we cannot rely on stationary and independent increments to simplify the expectations. Also, $\lambda_t, N_t$, and $B(N(t))$ are not pairwise independent and therefore none of the cross moments can be split. We do note however that since $B(t)$ is independent of $N_t$ and $\lambda_t$ by assumption, the values of $N_t$ and $\lambda_t$ only affect the occurrence rate of jumps of $B(N_t)$; they do not bias the values upwards or downwards. This then, implies the cross moments $\E(N_tB(N_t))$ and $\E((\lambda_t-v)B(N_t))$ are 0 as well. We may use equations \eqref{closedformEl2}--\eqref{closedformEN2} to calculate the generator. Assuming $\E(N_tB(N_t))=\E(\lambda_tB(N_t))=0$, we have every value for the generator except for $\E(B^2(N_t))$. We can calculate the generator termwise, using L'Hopital's rule where necessary:
\begin{align}
\lim_{t\downarrow 0}\frac{\E(\lambda_t-v)}{t}&=\alpha v,\\
\lim_{t\downarrow 0}\frac{\E(N_t)}{t}&=v,\\
\lim_{t\downarrow 0}\frac{\E(\lambda_tN_t)}{t}&=0,\\
\lim_{t\downarrow 0}\frac{\E(N_t^2)}{t}&=0,\\
\lim_{t\downarrow 0}\frac{\E((\lambda_t-v)^2)}{t}&=-2v^2(\alpha-\beta)+v(\alpha^2+2\beta v)-2v^2.
\end{align}
In order to calculate the above limits, we needed proper analytical expressions for the relevant expected values. These expressions have been determined explicitly in appendix \ref{explicitsol} by solving a simple system of three differential equations.
%%%

We are left with only one more unknown term. Since $\E(B(N_t))=0$, $\Var(B(N_t))=\E(B^2(N_t))$. 

We are able to calculate the second moment using conditioning. This will allow us to finalize the computation of the generator. We will complete this calculation with the formula: 
\[\Var(B(N_t))=\E[\Var(B(N_t)|N_t)]+\Var[\E(B(N_t)|N_t)].\]
We already know that the second term, $\E(B(N_t)|N_t)=0$ as a trivial fact of Brownian motions. We also see that $\E[\Var(B(N_t)|N_t)]=\E(N_t)$ by the properties of a Brownian motion. Thus, \[\Var(B(N_t))=\E[N_t].\] We have therefore found the generator:
\begin{align}
\mathcal{A}f(y)&=\lim_{t\downarrow 0}\frac{\E(A_2(Y_t)|Y_0=y)-f(Y_t)}{t}\\
&=\lim_{t\downarrow 0}\frac1t\bigg[\frac{\partial f(y)}{\partial l}\E(\lambda_t-v)+\frac{\partial f(y)}{\partial n}\E(N_t)+\frac12\frac{\partial^2f(y)}{\partial l^2}\E[(\lambda_t-v)^2]\nonumber\\
&\hspace{2.8cm}+\frac{\partial^2f(y)}{\partial l\partial n}\E[(\lambda_tN_t-vN_t]+\frac12\frac{\partial^2f(y)}{\partial b^2}\E[(B(N_t))^2]\bigg]\\
&=\frac{\partial f(y)}{\partial l}\alpha v+\frac{\partial f(y)}{\partial n}v+\frac12\frac{\partial^2f(y)}{\partial l^2}(-2v^2(\alpha-\beta)+v(\alpha^2+2\beta v)-2v^2)\nonumber\\
&\quad-\frac{\partial^2f(y)}{\partial l\partial n}v^2+\frac12\frac{\partial^2f(y)}{\partial b^2}v.
\end{align}

%%%%%%%%%%%%%%%%%%%%%%%%%%%%%%%%%%
\subsection{Ito Formula for $B^2(N_t)$}
\begin{theorem}
The Ito formula of a squared variance-Hawkes process is given by
\[dB^2(N_t)=2B(N_t)dB(N_t)+dN_t.\]
\end{theorem}
This theorem can be proven by writing $B^2(N_t)$ as $B(N_t)B(N_t)$ and applying the Ito product formula. Since the variance is known, the quadratic variation term simplifies readily.

We can simulate each side of the equation using the Euler--Maruyama approximation of the SDE. From there we can calculate the experimental distributions, observe the trajectories, and calculate the percentage errors between them. This will give us a reasonable measure of how closely the simulations of these processes agree. The left side of the SDE will be referred to as the {\emph{actual}} distribution or path and the right will be called the {\emph{conjectured}} distribution or path. These simulations will allow us to determine if these simulation methods are reliable, and also allow us to develop better intuition on how the variance of the variance Hawkes process behaves.

We chose the parameters to be $\alpha=600$, $\beta=800$, $v=5000$, $T=1$, RES$=2^{20}$. RES is the resolution of the simulation, i.e. the size of the vector containing the process. $\alpha$, $\beta$, and $T$ are taken be realistic values based on the calibrations done in a paper from Swishchuk et al.  \cite{swishchuk2019compound} and in the previous sections. Since $v$ is usually small relative to the duration of an average contract (say 1 month to 5 years), we have set it to be a sufficiently large value such that there are enough jumps in the process to permit some appropriate level of convergence in distributions. This has an added effect of causing large variance later in the simulations since variance scales linearly with $v$. Taking large $T$ leads to the same effect since the process's variance also scales approximately linearly with $T$.

Figures \ref{SDEApprox}--\ref{SDEApproxTrajError} show sixteen sample distributions, trajectories, and percentage errors. Note that these three figures were each generated using separate simulations, but use the same data between the actual and conjectured processes in each plot. 

The distributions in Figure \ref{SDEApprox} are very similar, however, the conjectured path appears to take mass away from the tall spikes near the origin when compared with the actual distribution. Otherwise, the two processes line up quite closely, even with some idiosyncrasies in their shapes. Due to the significant visual differences in each individual distribution, it appears that these distributions take an exceedingly long amount of time to converge to their proper shape.

Figure \ref{SDEApproxTraj} shows that on the macroscopic scale, these two processes look virtually identical. It is only when we zoom in very closely that errors start to emerge (see the bottom right panel). These errors are still large overall, but their behaviour is still remarkably similar. It is plausible that the errors are artifacts of the Euler--Maruyama discretization, which is known to harbor some inaccuracies. 

Figure \ref{SDEApproxTrajError} provides a slightly more objective look at the differences between these two functions by looking at their percentage error. Here, we see errors mostly between $-\%20$ and $\%20$ with occasional spikes in error exceeding \%$10^9$. These large spikes may be numerical artifacts as the processes quickly converge back to overall large, but relatively small errors. The \%200 error present in some of the process simulations exists from the start, implying that we are off by a constant number to make these processes agree in these cases. This is likely due to the discretization step in the simulations. When the Hawkes process is simulated, it returns a vector of arrival times. In order to discretize the process we must place this vector in a vector of size RES where each entry represents a timestep of size $\Delta t=RES^{-1}$. This becomes complicated since some time steps may include multiple jumps. This has the effect of essentially rounding up all jumps within an interval to the end of the closest time interval. Differences in when these jumps are recorded by the discretization algorithm are likely what lead to these large spikes. In most other simulations, the errors remain (except for random spikes) below \%20 error. The large error spikes do not appear to accumulate the error and do not typically cause the two processes to diverge. The bottom right panel is zoomed in to show the error on the smaller scale that is being hidden by the large spikes.

From these three figures,  it appears the conjectured path is largely of the correct form, but might be missing a correction term or some other small addition to agree precisely with the actual path.

\noindent{\bf Remark.} We present the following conjecture:
\begin{conj}\label{bntconj}
If B(t) is a standard Brownian motion and $N_t$ is a Hawkes process, then 
\begin{equation}
B(N_t)\disteq \sqrt{N_t} \mathcal{N}(0,1)
\end{equation}
where $\mathcal{N}(0,1)$ is a standard normal distribution.
\end{conj}
We can simulate the Hawkes process and a Brownian motion using the usual techniques. Taking their composition we obtain a simulation of $B(N_t)$. This allows us to use Monte-Carlo methods to compare the distribution of $B(N_t)$ to our theorized distribution $\sqrt{N_t} \mathcal{N}(0,1)$.

Figure \ref{approxgrid} demonstrates that regardless of parameter choices, it appears that this approximation is a close fit to the distribution of $B(N_t)$. There are a few plots which exhibit higher and steeper peaks than the predicted PDF, but regardless these are in the minority and the tail behaviour is matched in all plots. A standard normal distribution has been plotted in orange for reference.

\section{Conclusion}
In brief:
\begin{itemize}
\item We proposed a new tractable model framework called the variance-Hawkes process.
\item We calculated the generator and moments for the variance-Hawkes process.
\item We calculated the Ito formula for the variance-Hawkes process's variance.
\item The variance-Hawkes process fitted the log returns of fossil fuel futures.
\end{itemize}
In details, this paper presented a new tractable model framework, called the variance-Hawkes process, that exhibits volatility clustering with very few parameters. This model uses a new approach to spot price modeling using a discontinuous subordinator. The discontinuous subordinator necessitates a change in perspective on subordination from time change to tracking order arrival behaviours. This framework is flexible enough to be appended onto current models to instill them with clustering behavior. We demonstrated using a simple model and hand calibration that this model can be fit to 2018 and 2019 crude oil and natural gas front month futures spot prices. These fits agreed reasonably well to the experimental data, despite the simplicity of the model used. This indicated that a more sophisticated model may be worth pursuing for a better fit. We derived the generator of the triplet $(\lambda_t,N_t,B(N_t))$ using the properties of the Hawkes process and some properties of a standard Brownian motion. Finally, we calculated and analyzed simulations of the Ito formula for the variance of a Hawkes subordinated Brownian motion. These simulations agreed well with the Ito formula for the variance. Some discrepancies were present, highlighting inaccuracies with the simulation strategies used. These errors did not accumulate and were mostly consistent throughout the simulations. There were occasionally large spikes present in the error plots of the simulations, but these spikes were a temporary artifact of the simulation regime. Future work will include derivations for more complex models using Hawkes subordination such as those stated in section \ref{SubordAndCluster}. We will work towards further analysis of this process, proving Conjecture \ref{bntconj}, i.e.~its distributional properties, risk-neutral valuations, and derive calibration techniques for this new class of models. 

\section*{Acknowledgements}
The authors would like to thank NSERC for continuing support and the referees for their helpful comments.

\section*{Declarations of Interest}
The authors report no conflicts of interest. The authors alone are responsible for the content and writing of the paper.

%%%%%%%%%%%%%%%%%%%%%%%%%%%%%%%%%%
%\nocite{*}
\bibliographystyle{apalike}
\bibliography{BNtRefs}
%%%%%%%%%%%%%%%%%%%%%%%%%%%%%%%%%%
\appendix
\section{Supplemental Figures}
\subsection{2018 and 2019 natural gas and crude oil time series}
\begin{figure}[h!]
\centering
\subfloat{\scalebox{0.5}{\includegraphics[width=0.9\linewidth]{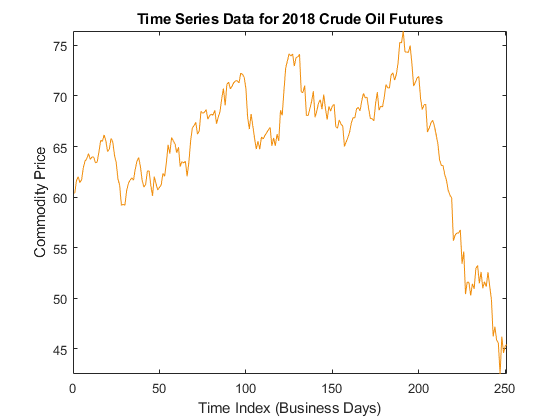}}}
\caption{2018 front-month WTI crude oil futures closing prices.}\label{TS1}%
\end{figure}

\begin{figure}[h!]
\centering
\subfloat{\scalebox{0.5}{\includegraphics[width=0.9\linewidth]{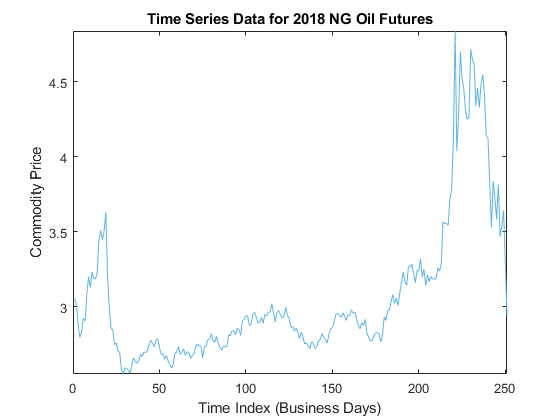}}}
\caption{2018 front-month NYMEX natural gas futures closing prices.}\label{TS2}%
\end{figure}

\begin{figure}[h!]
\centering
\subfloat{\scalebox{0.5}{\includegraphics[width=0.9\linewidth]{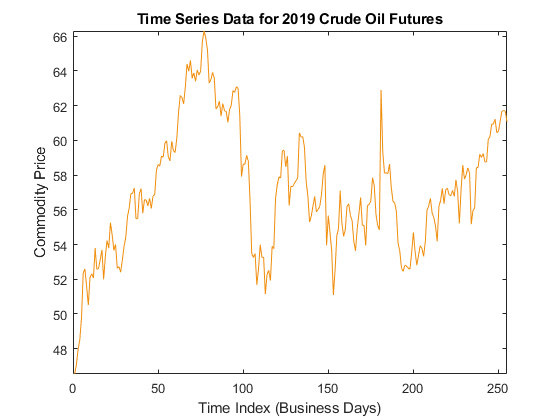}}}
\caption{2019 front-month WTI crude oil futures closing prices.}\label{TS3}%
\end{figure}

\begin{figure}[h!]
\centering
\subfloat{\scalebox{0.5}{\includegraphics[width=0.9\linewidth]{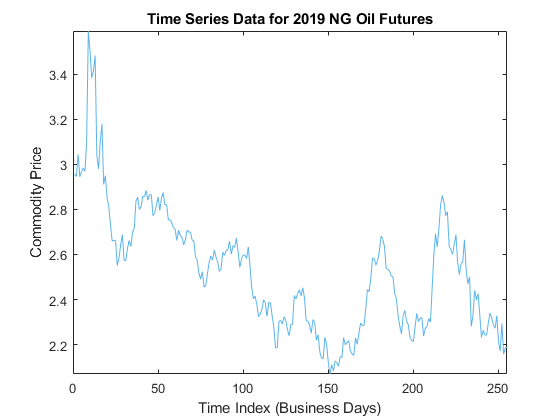}}}
\caption{2019 front-month NYMEX natural gas futures closing prices.}\label{TS4}%
\end{figure}

Figures \ref{TS1}--\ref{TS4} are the time series of one year of 2018 and 2019 front-month NYMEX natural gas and WTI crude oil futures. These datasets were obtained via Marketwatch.
\newpage
\subsection{Simulations for $B(N_t)$}
\begin{figure}[h!]
\centering
\subfloat{\hspace{-1cm}\scalebox{1}{\includegraphics[width=1.1\linewidth]{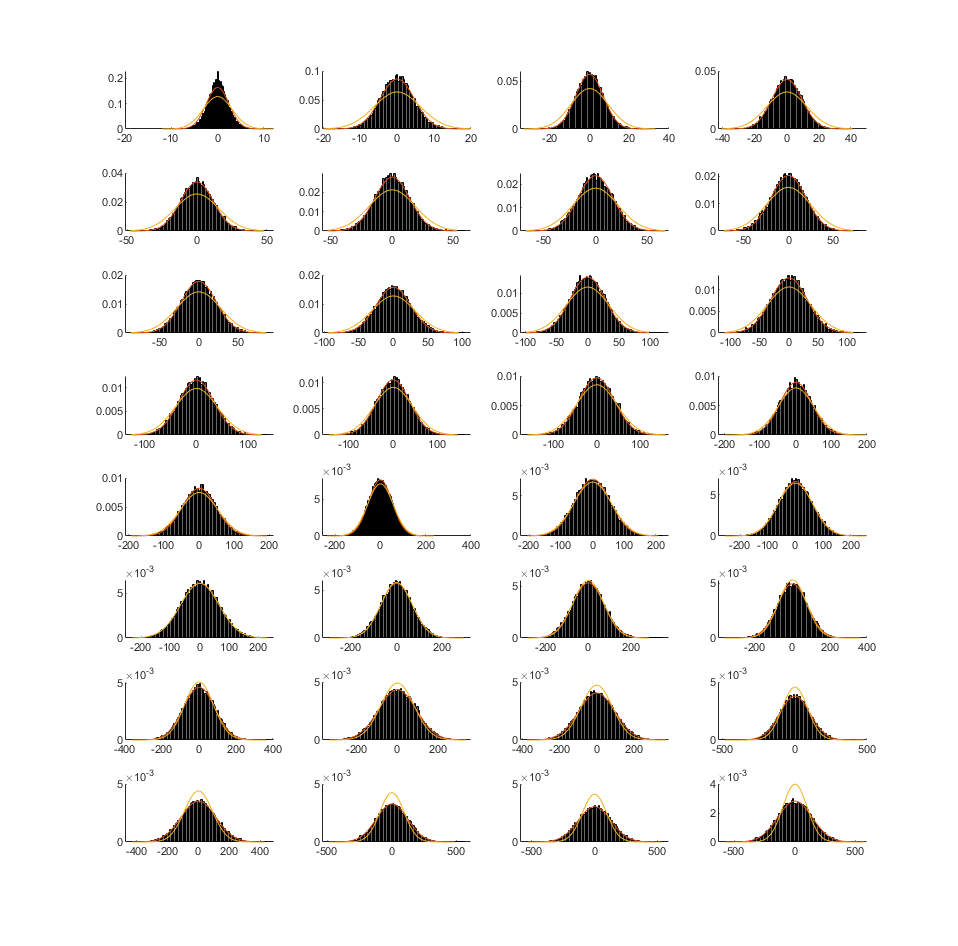}}}
\caption{Each plot is the distribution of $2^{14}$ simulations of $B(N_t)$ where the red line denotes the PDF of a normal distribution with mean 0 and variance $N_t$ and the orange line denotes the pdf of a corresponding Brownian motion for reference.}\label{approxgrid}%
\end{figure}

Figure \ref{approxgrid} shows simulated densities plotted with their predicted densities based on Conjecture \ref{bntconj}. The orange line is a standard normal distribution for reference.

\newpage

\begin{figure}[h!]
\centering
\subfloat{\scalebox{1}{\includegraphics[width=0.9\linewidth]{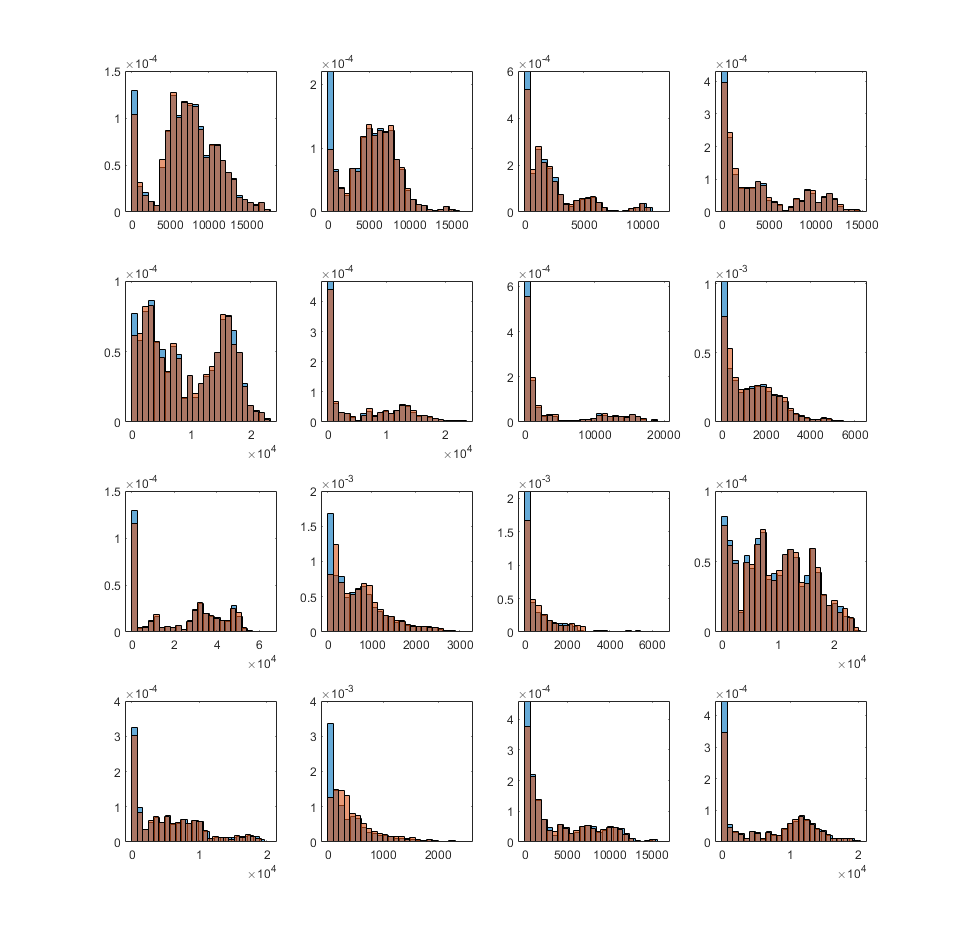}}}
\caption{Sixteen simulated PDFs of the processes $dB^2(N_t)$ (blue) and $2B(N_t)dB(N_t)+dNt$ (red).}\label{SDEApprox}%
\end{figure}
\newpage
\begin{figure}[h!]
\centering
\subfloat{\scalebox{1}{\includegraphics[width=0.9\linewidth]{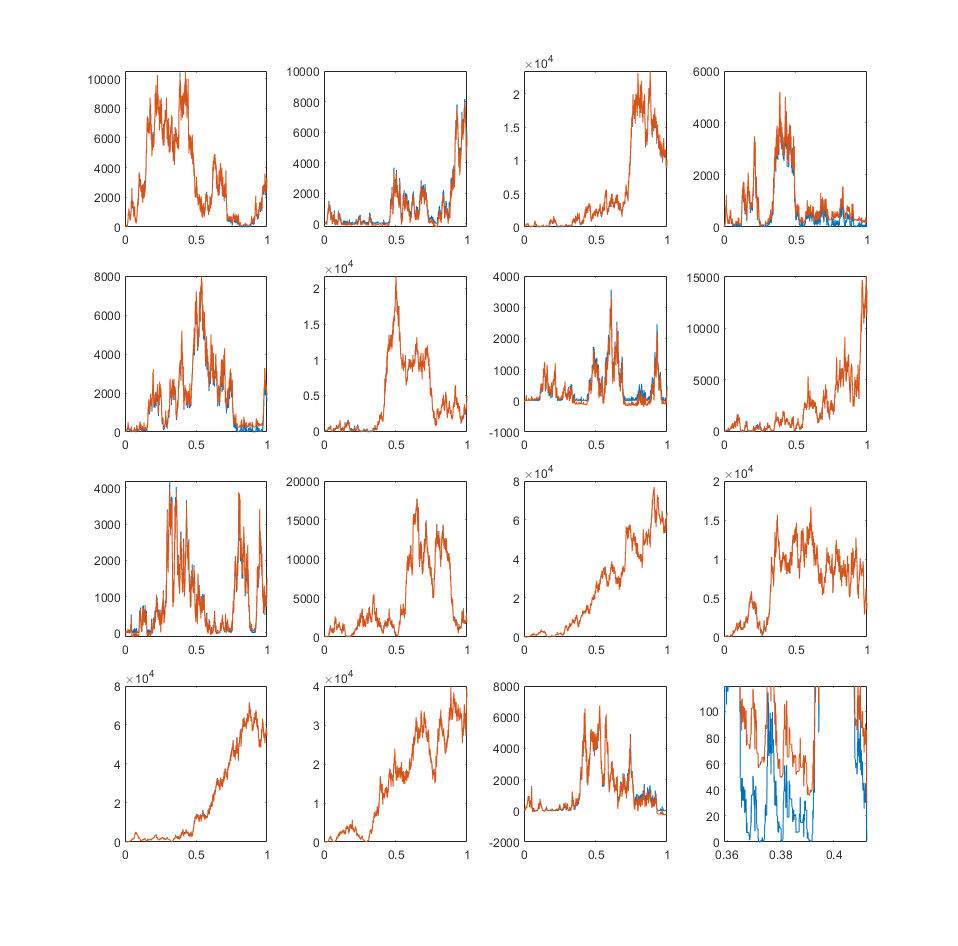}}}
\caption{Sixteen simulated trajectories of the processes $dB^2(N_t)$ (blue) and $2B(N_t)dB(N_t)+dN_t$ (red); the bottom rightmost figure has been zoomed in to highlight the differences between the processes more clearly.}\label{SDEApproxTraj}%
\end{figure}
\newpage
\begin{figure}[h!]
\centering
\subfloat{\scalebox{1}{\includegraphics[width=0.9\linewidth]{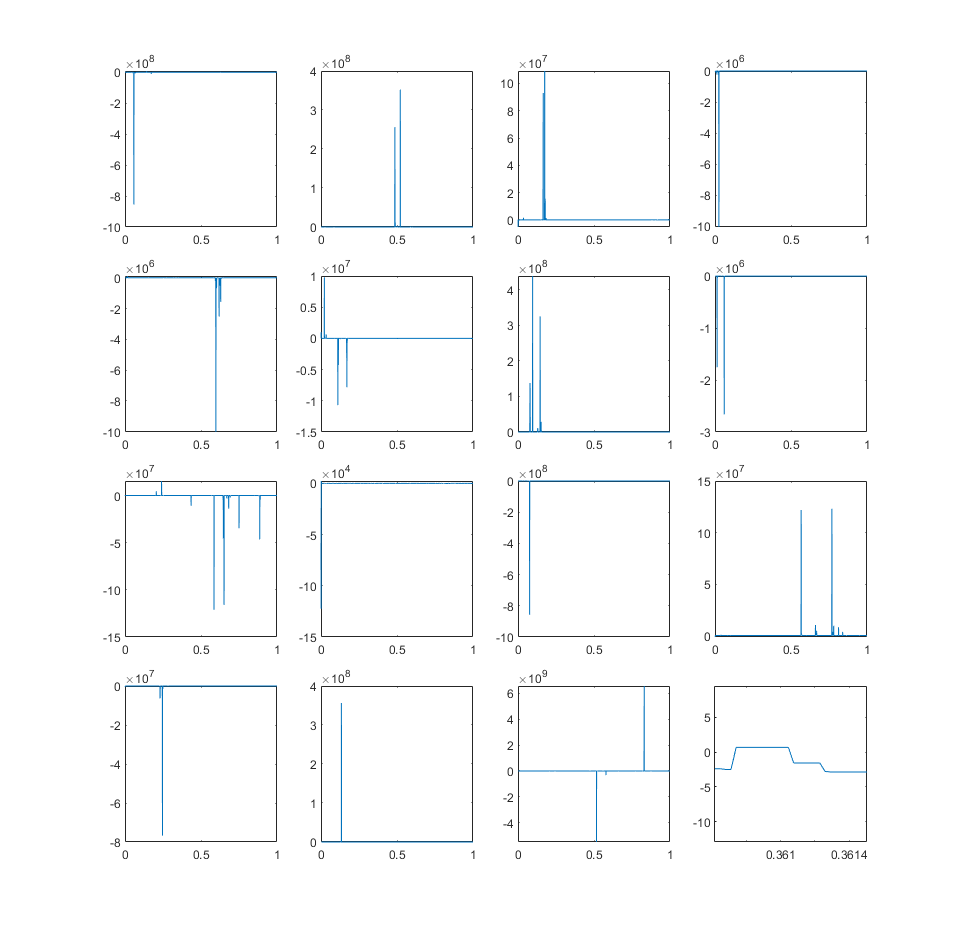}}}
\caption{Sixteen simulated trajectory percentage errors of the processes $dB^2(N_t)$ (blue) and $2B(N_t)dB(N_t)+dNt$ (red); the bottom rightmost figure has been zoomed in to highlight the smaller errors in the simulaitons.}\label{SDEApproxTrajError}%
\end{figure}
\clearpage

\section{Explicit expressions for the second and cross moments between $\lambda_t$ and $N_t$}\label{explicitsol}

We have three differential equations to solve which are obtained from equation \eqref{mom}. The first two solutions were given explicitly by \cite{cui2020elementary}. Since these expressions are large and unruly but ultimately simple exponentials, the only obstacle to integration is the successful management of the algebra. Therefore we can use MATLAB's symbolic computation package to obtain the closed-form solutions of $\E(N^2_t)$, $\E(\lambda_t^2)$, and $\E(\lambda_tN_t)$ easily. The second moment of $\lambda_t$ is given by
\begin{align}
\E(\lambda_t^2)&={\mathrm{e}}^{-t\,\left(-2\,\beta+2\,\alpha \right)}\,\left(v^2-A+B_t\right),\label{closedformEl2}
\end{align}
where 
\begin{align}
A=\frac{v\,\left(\alpha ^2+2\,\beta\,v\right)\,\left(\frac{\alpha }{-3\,\beta+3\,\alpha }-\frac{\beta}{-2\,\beta+2\,\alpha }\right)}{\alpha -\beta},
\end{align}
and
\begin{align}
B_t=\frac{v\,\left(\alpha ^2+2\,\beta\,v\right)\,\left(\frac{\alpha \,{\mathrm{e}}^{t\,\left(-3\,\beta+3\,\alpha \right)}}{-3\,\beta+3\,\alpha }-\frac{\beta\,{\mathrm{e}}^{t\,\left(-2\,\beta+2\,\alpha \right)}}{-2\,\beta+2\,\alpha }\right)}{\alpha -\beta}.
\end{align}
The cross moment is given by
\begin{align}
\E(\lambda_tN_t)=\frac{t^2\,{\mathrm{e}}^{-t\,\left(\alpha -\beta\right)}\,(C_t-D_t)}{36\,{\left(\alpha -\beta\right)}^2},
\end{align}
where
\begin{align}
C_t&=9\,\alpha ^2\,\beta+12\,\alpha ^2\,t+24\,\beta^2\,t-6\,\alpha ^3+18\,\alpha ^2\,{\mathrm{e}}^{3\,t\,\left(\alpha -\beta\right)}+6\,\alpha ^3\,{\mathrm{e}}^{3\,t\,\left(\alpha -\beta\right)}\nonumber\\
&\quad+18\,\beta^2\,{\mathrm{e}}^{2\,t\,\left(\alpha -\beta\right)}-9\,\alpha ^2\,\beta\,{\mathrm{e}}^{2\,t\,\left(\alpha -\beta\right)}-12\,\beta^2\,t\,{\mathrm{e}}^{2\,t\,\left(\alpha -\beta\right)},
\end{align}
and
\begin{align}
D_t&=32\,\alpha \,\beta\,t+12\,\beta^3\,t^2\,{\mathrm{e}}^{2\,t\,\left(\alpha -\beta\right)}-18\,\alpha \,\beta\,{\mathrm{e}}^{2\,t\,\left(\alpha -\beta\right)}-18\,\alpha \,\beta\,{\mathrm{e}}^{3\,t\,\left(\alpha -\beta\right)}\nonumber\\
&\quad-12\,\alpha \,\beta^2\,t^2\,{\mathrm{e}}^{2\,t\,\left(\alpha -\beta\right)}-12\,\alpha \,\beta\,t\,{\mathrm{e}}^{2\,t\,\left(\alpha -\beta\right)}+20\,\alpha \,\beta\,t\,{\mathrm{e}}^{3\,t\,\left(\alpha -\beta\right)}.
\end{align}
We are now able to solve the differential equation for $\E(N_t)$. The second moment of $N_t$ is given by
\begin{align}
\E(N^2_t)=\frac{t^3\,{\mathrm{e}}^{-t\,\left(\alpha -\beta\right)}\,(E_t-F_t)}{18\,{\left(\alpha -\beta\right)}^2}-t^2\,\left(\frac{\beta}{2\,\left(\alpha -\beta\right)}-\frac{\alpha \,{\mathrm{e}}^{t\,\left(\alpha -\beta\right)}}{2\,\left(\alpha -\beta\right)}\right),\label{closedformENt2}
\end{align}
where
\begin{align}
E_t&=9\,\alpha ^2\,\beta+12\,\alpha ^2\,t+24\,\beta^2\,t-6\,\alpha ^3+18\,\alpha ^2\,{\mathrm{e}}^{3\,t\,\left(\alpha -\beta\right)}+6\,\alpha ^3\,{\mathrm{e}}^{3\,t\,\left(\alpha -\beta\right)}\nonumber\\
&\quad+18\,\beta^2\,{\mathrm{e}}^{2\,t\,\left(\alpha -\beta\right)}-9\,\alpha ^2\,\beta\,{\mathrm{e}}^{2\,t\,\left(\alpha -\beta\right)}-12\,\beta^2\,t\,{\mathrm{e}}^{2\,t\,\left(\alpha -\beta\right)},
\end{align}
and
\begin{align}\label{closedformEN2}
F_t&=32\,\alpha \,\beta\,t+12\,\beta^3\,t^2\,{\mathrm{e}}^{2\,t\,\left(\alpha -\beta\right)}-18\,\alpha \,\beta\,{\mathrm{e}}^{2\,t\,\left(\alpha -\beta\right)}-18\,\alpha \,\beta\,{\mathrm{e}}^{3\,t\,\left(\alpha -\beta\right)}\nonumber\\
&\quad-12\,\alpha \,\beta^2\,t^2\,{\mathrm{e}}^{2\,t\,\left(\alpha -\beta\right)}-12\,\alpha \,\beta\,t\,{\mathrm{e}}^{2\,t\,\left(\alpha -\beta\right)}+20\,\alpha \,\beta\,t\,{\mathrm{e}}^{3\,t\,\left(\alpha -\beta\right)}.
\end{align}

%Extra stuff if needed
%\appendix

%\section{Figures}\label{appFig}

%\begin{figure}[h!]
%\centering
%\subfloat{\scalebox{1}{\includegraphics[width=0.9\linewidth]{2019CrudeTS}}}
%\caption{2019 front month crude oil futures closing prices.}\label{TS2}%
%\end{figure}

\end{document}